# Electrochemical deposition of ZnO hierarchical nanostructures from hydrogel coated electrodes

Shuxi Dai,[a,b] Yinyong Li, †[a] , Zuliang Du, *[b] and Kenneth R. Carter*[a]

**Abstract:** The electrochemical deposition of ZnO hierarchical nanostructures directly from PHEMA hydrogel coated electrodes has been successfully demonstrated. A variety of hierarchical ZnO nanostructures, including porous nanoflakes, nanosheets and nanopillar arrays were fabricated directly from the PHEMA hydrogel coated electrodes. Hybrid ZnO-hydrogel composite films were formed with low zinc concentration and short electrodeposition time. A dual-layer structure consisting of a ZnO/polymer and pure ZnO layer was obtained with zinc concentration above 0.01 M. SEM observations and XPS depth profiling were used to investigate ZnO nanostructure formation in the early electrodeposition process. A growth mechanism to understand the formation of ZnO/hydrogel hybrid hierarchical nanostructures was developed. The I-V characteristics of the ZnO-hydrogel composite films in dark and under ultraviolet (UV) illumination demonstrate potential applications in UV photodetection.

**Introduction**

ZnO is one of the most attractive oxide semiconductor materials with a wide band gap (3.4 eV) and large exciton binding energy (60 meV) for the promising applications in optoelectronic devices. [1] The fabrication of a variety of ZnO nanostructures, such as one dimensional nanowires/nanorods and two dimensional nanostructures have been extensively studied in the last two decades. [2, 3] Recently, hierarchical nanostructures have attracted considerable attention owing to their promising applications to nanodevices such as light-emitting diodes,[4] field-effect transistors,[5] chemical sensors,[6] solar cells, [7] etc. There have been many reports about the physical methods to fabricate hierarchical ZnO nanostructures including high temperature chemical vapor deposition (CVD),[8] thermal evaporation[9] and pulsed laser deposition (PLD).[10, 11] However, the high growth temperature limits the choice of substrates and needs expensive vacuum equipments.[12] It still remains a big challenge to develop simple and reliable low-temperature fabrication methods for ZnO hierarchical nanostructures with controlled morphology and crystal nature.

Hierarchical ZnO nanostructures can be synthesized at low-temperatures by various method including chemical bath deposition,[13] hydrothermal synthesis,[14] and eletrodeposition. [15] Among these solution based growth methods, electrochemical deposition (ECD) is a rapid and cost-effective approach for the fabrication of hierarchical ZnO nanostructures.[16] Peulon et al.[17] and Izaki et al.[18, 19] had performed the pioneering work in the field of electrodeposition of ZnO thin films and nanorods on ITO substrates. There have been a variety of reports on the electrochemical synthesis of ZnO nanostructures on various substrates, including GaN,[20] FTO, [21] Au/Si, [22] Zn foils[23]. However, there are only a few reports on the electrodeposition of ZnO nanostructures on the functional materials modified electrodes. Recently, Seong et al. prepared ZnO nanosheets and nanorods structures by electrodeposition on single-walled carbon nanotubes modified electrodes.[24] In particular, Ryan et al. had successfully electrodeposited ZnO nanostructures onto organic-semiconducting substrates comprising copper phthalocyanine and pentacene molecular thin films using a two-step electrochemical process. [25]

Hybrid organic–inorganic composite materials have attracted interest over the last decade.[26] A polymeric matrix not only provides a mechanical support for the functional inorganic materials but also add new interesting features to the hybrid materials.[27] Because of its water-swelling and biocompatible properties, poly(hydroxyethyl methacrylate) (PHEMA), a widely used hydrogel, has been commonly employed as polymer matrix materials for the fabrication of inorganic materials to get functional hybrid materials for many potential applications in photonics and biosensor.[28] The important advantage consists in combining useful properties of both constituting components, e.g. the flexibility and shaping versatility of PHEMA hydrogel and optoelectronic response of the inorganic component.[29] Many studies have been devoted to the preparation of PHEMA-based composite materials with organic HEMA monomers and inorganic precursors such as tetraethoxysilane (TEOS) or titanium-oxo through conventional sol-gel processes.[30] To our knowledge, electrodeposition of ZnO on the three-dimensional networks of PHEMA hydrogel and study of the growth mechanisms for the deposition of ZnO nanostructures on the hydrogel coated

† Co-first author

[a]  Dr. S. Dai, Y. Li, Prof. K. R. Carter
     Department of Polymer Science and Engineering, University of Massachusetts Amherst, Conte Center for Polymer Research, 120 Governors Drive, Amherst, Massachusetts 01003 (United States)
     E-mail: krcarter@polysci.umass.edu

[b]  Dr. S. Dai, Prof. Z. Du, zld@henu.edu.cn
     Key Laboratory for Special Functional Materials of Ministry of Education, Henan University, Kaifeng 475004 (P. R. China)

substrates have not yet been reported. It represents a new motif for the generation of inorganic/organic hybrid materials with useful electrical and optical properties.

In this study, ZnO hierarchical nanostructures were synthesized by a simple electrodeposition method directly from the PHEMA hydrogel coated electrodes (Scheme 1). This new method provides a simple and unique approach to fabricate hybrid polymer/ZnO composite films. Various hierarchical ZnO nanostructures were synthesized on the hydrogel coated electrodes. The growth mechanism is also discussed to understand the formation of ZnO/polymer hybrid hierarchical nanostructures. Hybrid UV photodetector devices based on Au/ZnO-hydrogel/ITO structures were fabricated. The expected UV photoresponse of the electrodeposited polymer/ZnO composite films was observed. This study provides a simple and unique approach to fabricate low-cost hybrid UV detectors.

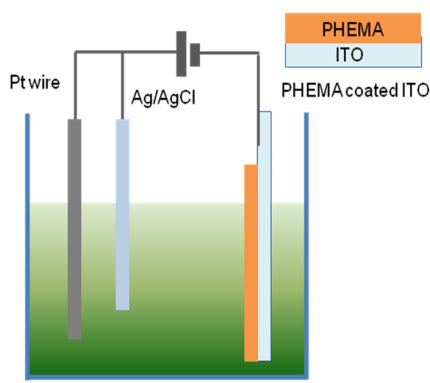

Scheme 1. Schematic diagram of the three-electrode setup for the electrochemical deposition of ZnO hierarchical nanostructures on PHEMA hydrogel coated electrodes.

### Results and Discussion

The morphology of the ZnO hierarchical nanostructures fabricated on hydrogel films via electrochemical deposition was observed by SEM. Figure 1 shows the typical top view and cross-sectional SEM images of typical ZnO nanostructures cathodically deposited with constant electrochemical parameters ($Zn^{2+}$ = 0.1 M, T = 70 $^{\circ}$C, t = 1000 s, E = -1.1 V) on PHEMA hydrogel films with different thickness. As shown in Figure 1a, large-scale nanoflowers structures were obtained on the 500-nm-thick PHEMA hydrogel films after 1000 s electrodeposition. Each flower consisted of 3 – 5 petal-like structures with a thickness less than 50 nm and length ranging from 200 nm to 1000 nm. Figure 1c presents the crystalline ZnO film with ridge-like surface morphology deposited on the 250-nm-thick PHEMA hydrogel films. Cross-section images of Figure 1b and 1d clearly show a dual layer structure of electrodeposited ZnO films. The bottom layer corresponded to PHEMA/ZnO composite film and the top layer corresponded to a pure layer of crystalline ZnO hierarchical nanostructures. For the ZnO nanostructures deposited on the 500-nm-thick PHEMA hydrogel (Figure 1b), the top layer of crystalline ZnO film was 3 μm thick. A 1.5 μm thickness of bottom layer of PHEMA/ZnO composite film is about 3 times larger than that of the original 500-nm dry PHEMA hydrogel. Swelling measurements of crosslinked PHEMA hydrogels were performed under the same electrodeposition condition. The crosslinked PEMA hydrogel exhibited a swelling behavior with an average swelling ratio of about 1.21 measured after immersed in the zinc nitrate solution for 30 min at 70 $^{\circ}$C. It indicated the increased thickness of ZnO/PHEMA composite films compared to that of a dry PHEMA hydrogel film is largely influenced by the formation and growth of ZnO nanocrystals inside the PHEMA hydrogel three dimensional networks.

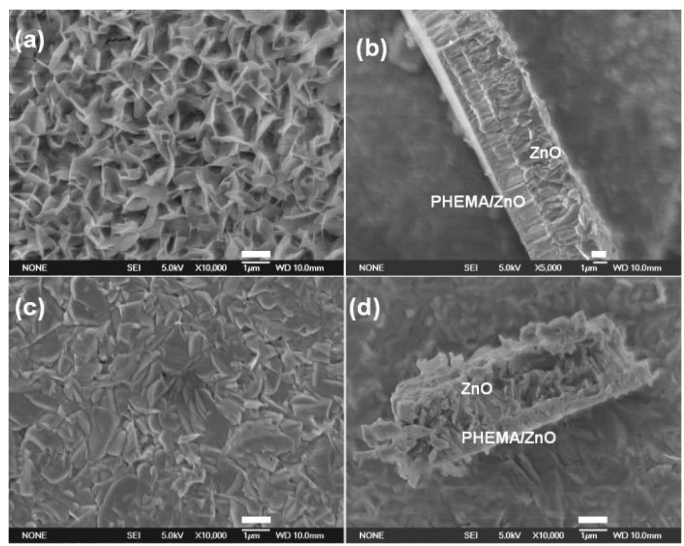

Figure 1. Top-view and cross-sectional SEM images of ZnO nanostructures electrodeposited at 70 $^{\circ}$C in 0.1 M zinc nitrate solution under -1.1 V for 1000 s on PHEMA hydrogel films with thickness of (a, b) 500 nm and (c, d) 250 nm. Scale bar = 1μm.

Our experiments reveal that the morphology and crystal nature of the electrodeposited ZnO hierarchical structures on PHEMA hydrogel coated ITO could be tuned by adjusting the deposition conditions, such as concentration of zinc ions in the electrolyte, applied potential and eletrodeposition time. Figure 2 presents the SEM images of ZnO nanostructures cathodically deposited on crosslinked PHEMA coated ITO with increasing zinc nitrate solutions from 0.01 M to 0.2 M with other constant electrochemical parameters (E = -1.1 V, T = 70 $^{\circ}$C, t = 1000 s, hydrogel film thickness = 500 nm). Figure 2a shows the surface morphology of ZnO films deposited on the PHEMA coated ITO in the 0.01 M zinc nitrate solution. A relatively flat film composed of particles with a size ranging from 100 to 300 nm and some pieces of thin nanoflakes with a thickness of 50 nm can be observed.

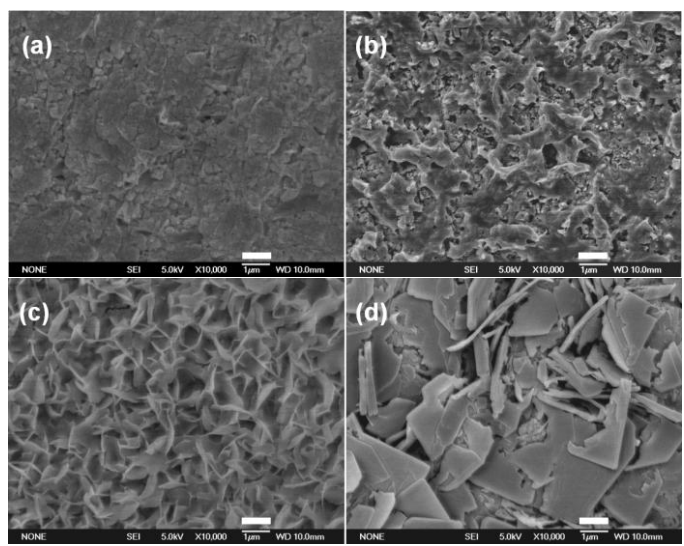

Figure 2. Top view SEM images of ZnO films electrodeposited on PHEMA hydrogel coated ITO substrates with constant electrochemical parameters (E = -1.1 V, T = 70 $^{\circ}$C, t = 1000 s, hydrogel film thickness = 500 nm) at different $Zn(NO_3)_2$ concentrations of (a) 0.01M, (b) 0.05M, (c) 0.1M and (d) 0.2M. Scale bar = 1μm.

As the electrolyte concentration increased to 0.05 M, irregular porous surfaces appeared on the PHEMA/ITO electrode. Figure 2b clearly shows the porous network structures of nanoflakes with thickness of 50 - 100 nm. The diameter of large pores in the porous network is about 400 - 800 nm. Nanoflower structures with thickness of 50 nm and length less than 1 μm were obtained after 1000 s electrodeposition in 0.1 M electrolyte (Figure 2c). Figure 2d presents typical morphology of sheet-like ZnO nanostructures obtained in 0.2 M zinc ions solution. The nanosheets had unordered orientations and smooth surfaces with thickness less than 100 nm and width of 2 - 3 μm.

Figure 3 shows the XRD patterns of the electrodeposited ZnO films on the PHEMA hydrogel coated ITO substrates at 70 °C under -1.1 V for 1000 s at different Zn ions concentrations range from 0.001 M to 0.1 M. The zinc ions concentration was found to influence the growth rate and the crystal nature of the films. A higher concentration of zinc ions had resulted in a faster growth rate. There is no obvious diffraction peak of ZnO for the film electrodeposited in the 0.001 M for 1000 s (Pattern b). Crystalline ZnO was formed only when the zinc ion concentration increased to 0.01 M. For XRD pattern c to e, the XRD diffraction peaks can be indexed as hexagonal wurtzite structured ZnO, which is in good agreement with the literature values (JCPDS card, No. 36-1451)[16] except the peaks marked with an asterisk symbol that result from the ITO substrate (JCPDS, No. 06-0416). The ZnO film deposited in the zinc nitrate solution above 0.05 M exhibits good crystallinity and no signal corresponding to Zn metal or other oxides is found. The presence of (100), (002) and (101) peaks indicated the random orientation of the ZnO films, which correspond to the SEM observation in Figure 2.

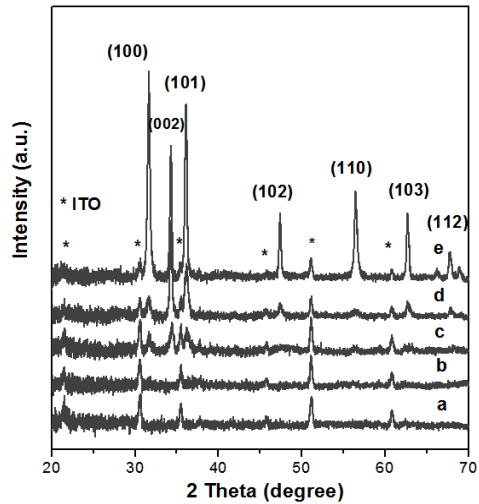

Figure 3. XRD patterns of (a) ITO substrate and ZnO films electrodeposited on PHEMA hydrogel coated ITO at 70 °C under -1.1 V for 1000 s at different $Zn(NO_3)_2$ concentrations of (b) 0.001M, (c) 0.01M (d) 0.05M, (e) 0.1M.

The electrochemical reaction is also determined by the electrodeposition potential.[31] Figure 4 shows the SEM images of ZnO nanostructures cathodically deposited on PHEMA coated ITO with constant electrochemical parameters ($Zn^{2+}$ = 0.01 M, T = 70 °C, t = 1000 s, PHEMA film thickness = 500 nm) under different applied potentials of -1.0 V, -1.1 V and -1.2 V (versus Ag/AgCl). Figure 4a shows the low magnification and high magnification SEM images (insert image in Figure 4a) of the electrodeposited ZnO films under -1.0 V with flat and compact surfaces. The morphology of ZnO films deposited under -1.1 V shows a porous nanoflake structures (Figure 4c). The ZnO film electrodeposited under potentiostatic condition of −1.2 V displays morphology of submicron pillar arrays. The diameters of the hexagonal faced pillars are in the range of 500 - 800 nm. The submicron pillar arrays show a relative preferential c-axis perpendicular to the substrate. These observations are in accordance with the presence of a strong (002) peak in XRD data.

According to the top-view SEM images (Figure 4a, 4c and 4e), the surface of electrodeposited ZnO films becomes more rough with the increased negative applied potential. The morphology changes from a flat surface to porous nanoflakes structures and then to submicron pillar arrays. Cross-section images of Figure 4b, 4d and 4f clearly show the dual layer structures of electrodeposited ZnO films. The PHEMA/ZnO composite layers deposited under different applied potentials of -1.0 V, -1.1 V and -1.2 V had a thickness of 1.2 - 1.5 μm The top crystalline ZnO layers showed a thickness of 0.5 μm, 1.0 μm and 1.5 μm for the applied potentials of -1.0 V, -1.1 V and -1.2 V. It indicated that the applied potentials exhibited more influence on the morphology and structures of top layer of ZnO films.

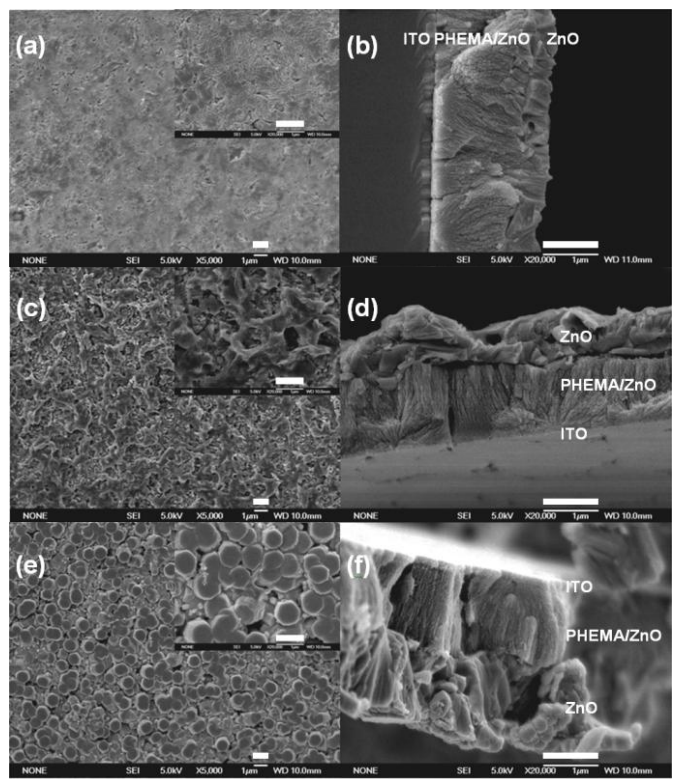

Figure 4. Top-view and cross-section SEM images of ZnO films electrodeposited at 70 °C for 1000 s in 0.05 M $Zn(NO_3)_2$ on 500-nm-thick PHEMA hydrogel coated ITO substrates under different applied potentials of (a, b) -1.0 V, (c, d) -1.1 V and (e, f) -1.2 V. Scale bar = 1μm.

The deposition of ZnO in the early stage generated large amounts of ZnO nanocrystals inside of the hydrogel networks and then filling in the bottom PHEMA layer. The former stage creates a compact bottom ZnO/PHEMA layer, while the following growth of ZnO nanocrystals out of the compact layer results in the formation of hierarchical nanostructures of porous flakes and pillars structures. High negative voltage can liberate more hydroxide ions and zinc ions in electrolyte readily diffused to or adsorbed on the cathode surface due to strong electric field intensity which catalyzed the electrodeposition process.[31] When high electrodeposition voltages of -1.2 V were applied, nanocrystals stacking along one preferential direction may dominate the ZnO deposition, thus leading to the appearance of ZnO pillars.

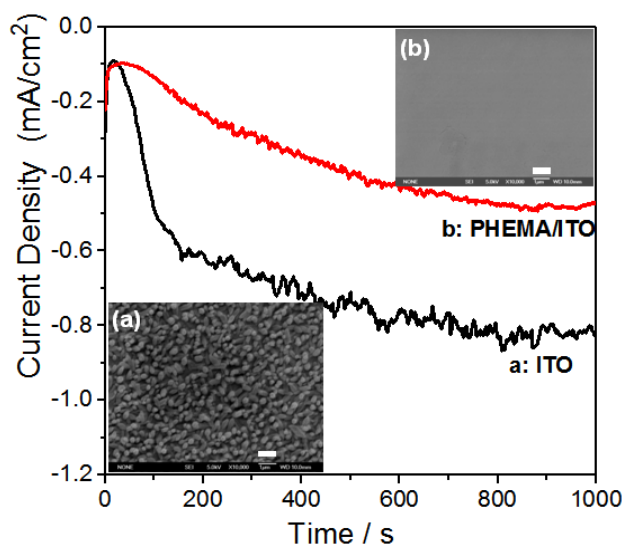

Figure 5. Variation curves of the cathodic current density as the function of the deposition time for ECD of ZnO films at 70 °C under -1.1 V from 0.001 M Zn(NO$_3$)$_2$ solution on (a) ITO and (b) 500-nm-thick PHEMA coated ITO substrates. Insert images shows the top-view SEM images of the ZnO nanostructures deposited on (a) ITO and (b) 500-nm-thick crosslinked PHEMA coated ITO substrates. Scale bar = 1μm.

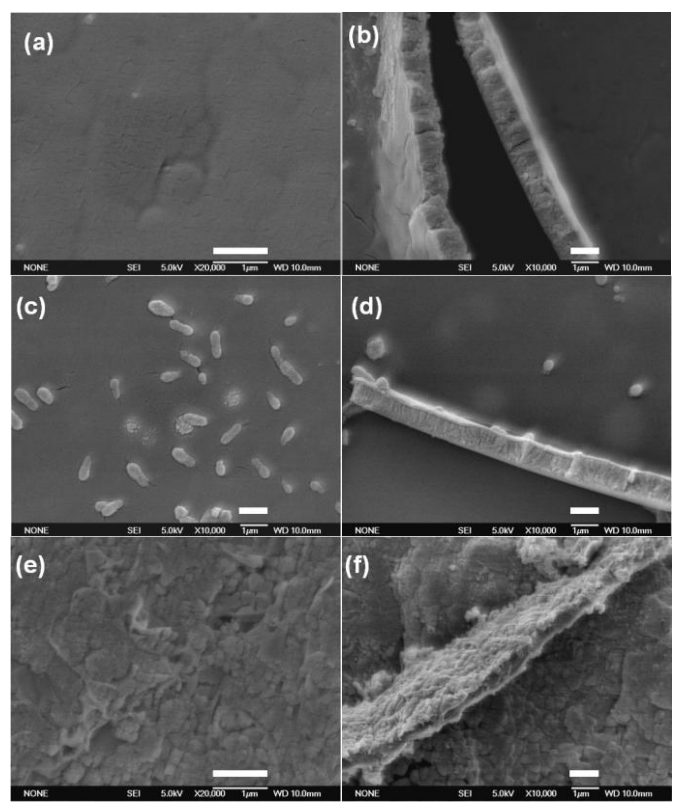

Figure 6. Top-view and cross-section SEM images of ZnO films electrodeposited at 70 °C under -1.1 V on 500-nm-thick PHEMA hydrogel coated ITO substrates in 0.01 M Zn(NO3)2 solutions for (a, b) 400 s, (c, d) 600 s and (e, f) 1000 s. Scale bar = 1μm.

Electrodeposition of ZnO is based on the generation of OH$^-$ ions at the surface of working electrode by electrochemical cathodic reduction of precursors such as O$_2$, NO$_3$- and H$_2$O$_2$ in zinc ions aqueous solution.[32] In our experiments, ZnO films were cathodically deposited from zinc nitrate solutions. The zinc nitrate solution can act as both the zinc and oxygen precursor. The general scheme of eletrodeposition of ZnO from aqueous zinc nitrate solution is supposed as follows (Eqs. (1) – (4)):

$$Zn(NO_3)_2 \rightarrow Zn^{2+} + 2NO_3^- \quad (1)$$

$$NO_3^- + H_2O + 2e^- \rightarrow NO_2^- + 2OH^- \quad (2)$$

$$Zn^{2+} + 2OH^- \rightarrow Zn(OH)_2 \quad (3)$$

$$Zn(OH)_2 \rightarrow ZnO + H_2O \quad (4)$$

Cathodic electrochemical reduction of nitrate to nitrite is catalyzed by Zn$^{2+}$ ions that are adsorbed on the surface of the cathode and liberates hydroxide ions, as in Eqs. (1). Then, zinc ions precipitate with the hydroxyl anions, resulting in the formation of zinc hydroxide. Subsequently, zinc hydroxide spontaneously dehydrates into zinc oxide at a slightly elevated temperature of about 60°C.[33] Finally, these series of multi-step reactions can be summarized by Eq. (5).

$$Zn^{2+} + NO_3^- + 2e^- \rightarrow ZnO + NO_2^- \quad (5)$$

The water adsorption and swelling properties make crosslinked PHEMA hydrogel a good candidate for the modification of working electrodes. Figure 5 presents the typical variation curves of the cathodic current density as the function of the deposition time for ECD of ZnO films on ITO and PHEMA hydrogel coated ITO substrates at 70 °C under a constant deposition potential of -1.1 V from 0.001 M Zn(NO$_3$)$_2$ solution. Figure 6 curve a represents the variation of current density with time for ECD of ZnO films on ITO substrate. It shows a rapid increase of current density to a value of -0.5 mA/cm$^2$ in the first 100 s, corresponding to the nucleation stage of ZnO crystallites. Then after 400 s the stable current density corresponds to growth of ZnO nanostructures. The insert SEM image of Figure 5a shows a clear morphology of ZnO nanorod arrays grown on the ITO substrate after 1000 s electrodeposition. Curve b in Figure 6 shows a nearly linear increase of current density until 600 s then the current density kept stable at a value of -0.5 mA/cm$^2$. The lower current density of the sample deposited on PHEMA hydrogel in curve b shows the ZnO crystallization process is slower than that on pure ITO substrates. Even the current density value at 1000 s for PHEMA coated substrate is lower than the value of 100 s for ITO substrate. It indicates the ZnO crystallites grown inside of the PHEMA hydrogel film is still in the nucleation stage. The insert SEM image of Figure 5b shows a smooth and compact surface with no obvious ZnO structures found on the ZnO/PHEMA composite films similar to morphology of the pure PHEMA film before electrodeposition.

The effect of substrates on the current density and surface morphology of electrodeposited ZnO films show that there is a different formation mechanism of ZnO nanostructures on the PHEMA coated electrode. The current density increases greatly on the ITO substrate in the first 100 s due to the fast reduction rate of nitrate ions to OH ions on the total ITO electrode surface. Then Zn$^{2+}$ ions in the vicinity of working electrode react with OH$^-$ ions, leading to the fast precipitation of ZnO on the ITO electrode surface and then following rapid growth of ZnO nanorod arrays. For PHEMA coated ITO electrode, as the concentration of zinc nitrate decreased to only 0.001 M, the zinc ions diffusion in PHEMA hydrogel was limited. Only a small amount of zinc ions could arrive at the electrode surface. The electrochemical reduced OH$^-$ ions would first interact with the Zn ions that have diffused near the ITO surface. Similarly, the hydroxyl ions formed at the electrode surface also diffused into the PHEMA network and interacted with the zinc ions. The carbonyl groups and free hydroxyl groups present in the hydrogel give a number of sites where both the Zn and OH$^-$ can significantly interact, slowing diffusion and reactivity. The ZnO nucleation starts either at the ITO surface or within the PHEMA network near the ITO substrate. Accordingly, the growth of ZnO nanostructures is relatively slow due to the interaction of zinc and OH$^-$ ions in the PHEMA hydrogel films.

In order to investigate the growth process of ZnO/PHEMA hybrid films, a series of samples were obtained by electrodeposited at 70 $^o$C under -1.1 V on 500-nm-thick crosslinked PHEMA hydrogel coated ITO substrates in 0.01 M Zn(NO$_3$)$_2$ solutions for 400 s, 600 s and 1000 s. For comparison, some electrodeposited samples were calcined at 500 $^o$C for 2 h in air to remove the polymer phase. Figure 6 shows the top-view and cross-section SEM images of electrodeposited ZnO/PHEMA films and Figure S1 present the surface morphologies of these ZnO films after calcination. As shown in Figure 6a and 6b, after eletrodeposition for 400s, the single layer of PHEMA/ZnO composite film had a flat surface morphology. It is still in the nucleation stage of the ZnO nanoparticles within the PHEMA networks. Figure S1a and S1b show the morphology and structures of 400 s deposited ZnO films after 2h calcination. Irregular micro-scale cracks were found on the surface of porous ZnO film composed of dispersed nanoparticles. For the samples obtained after 600 s electrodeposition, worm-like ZnO structures were found with width of 300 - 500 nm protruded out of the compact PHEMA/ZnO layer (Figure 6 c-d). It can be seen that the ZnO film consisted of vertically connected nanocrystals with the worm-like structures on top of the film. The ZnO film electrodeposited for 1000 s shows a rough surface consisted of particles and flakes (Figure 6 e-f). After calcination, the ZnO films can be distinguished as dual layer structure (Figure S1 e-f). It indicates that by increasing the deposition time, the ZnO deposition appears to follow the pathway of filling in the bottom PHEMA layer and growing out of the PHEMA/ZnO composite layer and forming a pure ZnO top layer by nanocrystals stacking.

XPS measurements were performed to investigate the elements distribution in the ZnO/PHEMA hybrid films. Figure S2 shows XPS survey spectra obtained for above mentioned ZnO/PHEMA composite films electrodeposited at the growth stages of 400 s and 1000 s. Zn2p, C1s, O1s and Si 1s were observed in the 400 s electrodeposited ZnO/PHEMA films and the peak positions of Zn2p3/2, Zn2p1/2 and O1s, C 1s are in good agreement with the reported values.[34],[35] The XPS survey spectra for 1000 s ZnO film shows prominent Zn 2p and O 1s features. The intensity of the Zn 2p peaks here are stronger than those of 400 s ZnO film. From the SEM image in Figure 6e and Figure S1e, we know that the strong signal for 1000 s sample is coming from the top layer of ZnO crystalline nanostructures. The minor carbon (C 1s) feature is also observed and is due to the surface contaminants arising from the sample collection and handling.

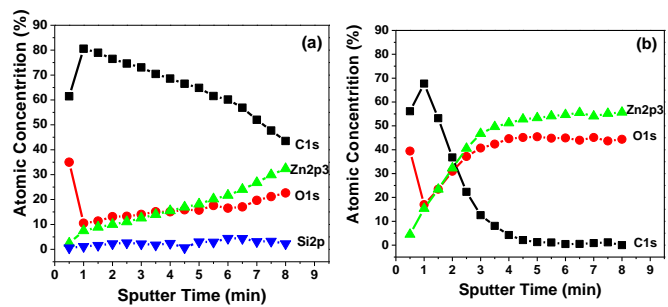

Figure 7. Composition depth profiles obtained from XPS analysis for the ZnO/PHEMA composite films electrodeposited at 70 $^o$C under -1.1 V on PHEMA hydrogel coated ITO substrates in 0.01 M Zn(NO$_3$)$_2$ solution for (a) 400 s and (b)1000 s.

XPS depth profiles measurements were simultaneously performed to get information on the distribution and the chemical state of zinc. Electrodeposited ZnO/PHEMA films were sequentially etched by Ar$^+$ ions for every 30 s and 480 s in total. XPS spectra were recorded after each sputtering step. The relative atomic concentration of zinc, carbon, oxygen and silicon was evaluated from the Zn 2p3, O 1s, C 1s, and Si 2s core level XPS spectra. In 3d spectra were also recorded after 480 s sputtering for all samples and no signal from the ITO substrate were detected by XPS. For the ZnO/PHEMA composite film electrodeposited for 400 s, a continuous increase of Zn and O concentration were observed while the carbon concentration gradually decreased with sputtering time (Figure 7a). The trace mount of Si corresponds to the silane crosslinking agent in the PHEMA networks. Also noted is the presence of carbon on the surface of ZnO film deposited for 1000 s (Figure 7b). The carbon concentration sharply decreased with increasing sputtering time and dropped below the detection limit after Ar+ ion sputtering for 240 s. As noted earlier, this carbon is due to surface contamination during sample handling in air. As indicated in Figure 7b, significant amounts of Zn ions were present after 120 s sputtering of outer contamination layer. After 300 s sputtering, there is only Zn and O left suggesting a deposited layer of zinc oxide. The Zn/O ratio was 1.2 after Ar+ ion sputtering for 180 s and was 1.19 after 480 s sputtering. This indicated that the composition of the ZnO film is not completely stoichiometric with oxygen deficiency.

From above results and discussion, we proposed a possible formation and growth mechanism of ZnO hierarchical nanostructures electrodeposited on the PHEMA hydrogel coated electrode. Figure 8 shows the schematic diagram for the formation and growth process. Firstly, the PHEMA hydrogel film on the substrate swells while immersed into the zinc nitrate electrolyte. The water permeable nature of PHEMA hydrogel film enables Zn$^{2+}$ and NO$_3^-$ ions diffuse into the interior of PHEMA hydrogel film. The positively charged zinc cation is distributed throughout and loosely bound with the carbonyl and hydroxyl groups within the PHEMA networks due to electrostatic interaction.

After the potential was applied, the zinc ions precipitated with the electrochemical reduced hydroxyl anions and then zinc hydroxide precipitations form. Subsequently, zinc hydroxide converts to zinc oxide at the processing temperature (70 $^o$C). The resulting ZnO nanoparticles are well dispersed in the PHEMA hydrogel three dimensional matrix. The confinement within the polymer matrix may prevent the ZnO nanoparticles from grossly aggregating. With continued electrodeposition, the ZnO nanoparticles continue to grow within the swollen PHEMA hydrogel and form ZnO nanocrystals. The ZnO nanocrystals eventually first fill the free space of the PHEMA hydrogel and continue to grow out of the PHEMA/ZnO composite film into the bulk electrolytes. Without the limitation of zinc ions concentration in the hydrogel, the emerging ZnO films grow fast and freely to yield a variety of different hierarchical nanostructures according to local electrochemical parameters.

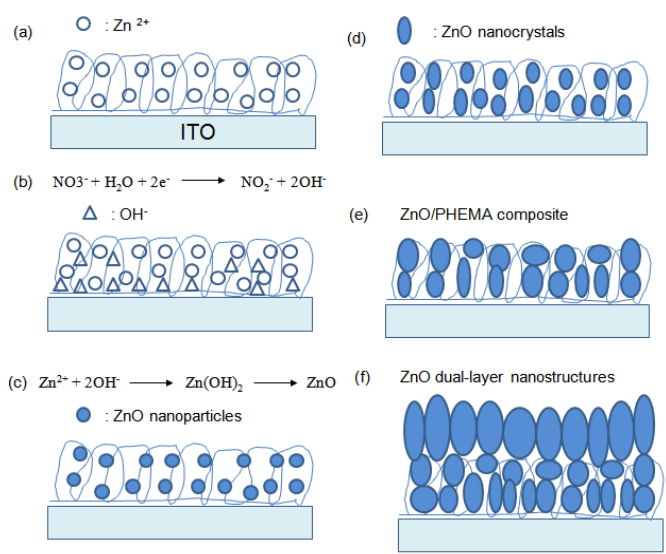

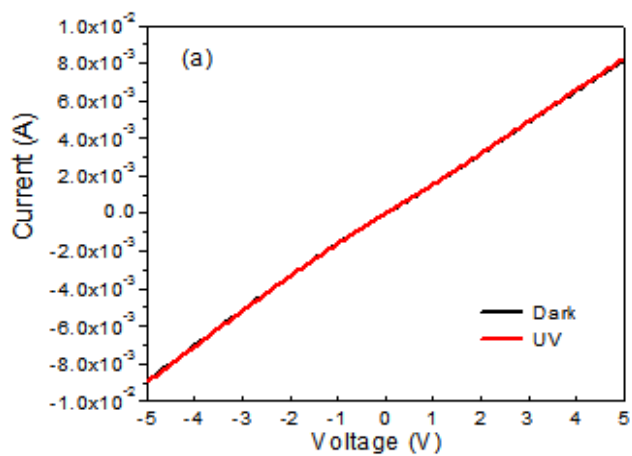

Figure 8. Schematic diagram for the fabrication of ZnO nanostructures on PHEMA coated ITO by electrochemical deposition. (a) adsorption of Zn ions, (b) cathodic electroreduction of hydroxide ions, (c) formation of zinc hydroxide and then zinc oxide, (d) growth of ZnO nanocrystals, (e) ZnO/PHEMA composite film and (f) ZnO dual-layer nanostructures.

Figure 9 shows the typical I-V characteristics of electrodeposited ZnO films in 0.01 M $Zn(NO_3)_2$ solution for 1000 s measured in the dark and under UV illumination. A linear curve under both forward and reverse bias was obtained in dark and UV illumination (Fig. 9a) for the ZnO film electrodeposited on pure ITO substrate. The linearity of the dark I-V curve indicates an ohmic behavior of the contacts. Under a 5 V applied bias, the dark current value is about 8 mA. The overlapping IV curves in the dark and under UV illumination (Fig.9a) indicate that the crystalline ZnO layer on ITO does not show UV sensitivity due to its initially high conductivity in dark state. Fig. 9b shows the I–V characteristics of devices based on electrodeposited ZnO/hydrogel composite films measured in the dark and under UV exposure. The dark current for a bias voltage of 5 V is 1.4 μA. The photocurrent upon exposure to 365 nm UV light is 72 μA at 5 V bias, which is significantly higher than the dark current, which represents an approximately 50x increase in current. When compared to the ZnO film on pure ITO, the higher surface area to volume ratio of the ZnO nanoparticles confined within the polymer matrix provides efficient absorption sites of oxygen molecules resulting in low conductivity in the dark. The enhanced photocurrent is associated with a photodesorption of loosely bound oxygen induced by UV light from the ZnO nanoparticles surfaces, thus removing electron traps and increasing the free carrier density.[36] The electrodeposited ZnO/hydrogel film show significant photocurrent generation and potential to be applied to a low-cost prototype UV detector. The influence of ZnO electrodeposition parameters on photoelectric performance of the ZnO/hydrogel based UV photodetector is under investigation.

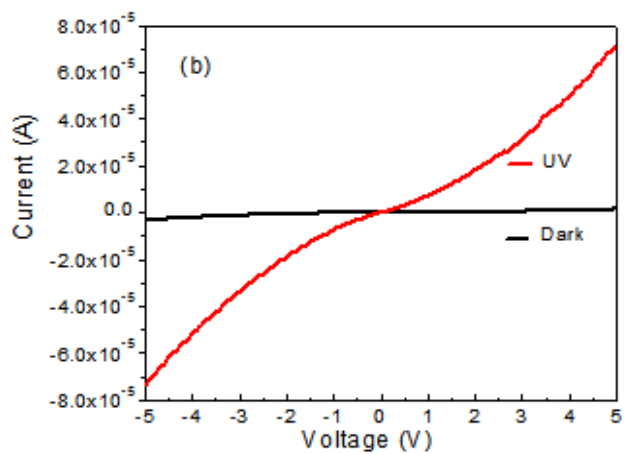

Figure 9. Dark and UV illuminated I-V Characteristics of ZnO films electrodeposited on (a) ITO and (b) hydrogel coated ITO substrates in 0.01 M Zn(NO3)2 solution for 1000 s.

### Conclusion

In summary, hierarchical ZnO nanostructures were obtained by direct electrochemical deposition on PHEMA hydrogel coated electrodes. Various morphologies of porous nanoflakes, nanosheets and nanopillar arrays were synthesized on the PHEMA hydrogel coated electrodes. Several parameters including the PHEMA hydrogel coating, concentration of zinc nitrate electrolyte, applied potentials and electrodeposition time, are all important in controlling the morphology and crystal nature of the hierarchical nanostructures. The photoelectrical measurements demonstrate that the device based on ZnO/hydrogel composite films showed a high sensitivity of UV light. This electrodeposition method using hydrogel-coated electrode represents a new motif for the preparation of inorganic/organic hybrid materials for optoelectronic applications. It can be expanded to other functional semiconductor materials and hydrogel systems for more potential applications.

### Experimental Section

*Materials*

All organic reagents and solvents were purchased from Sigma-Aldrich and used without further purification unless otherwise stated. Electrolyte solutions were prepared by dissolving zinc nitrate hexahydrate ($Zn(NO_3)_2 \cdot 6H_2O$), Aldrich, 99%) at concentrations ranging from 0.0001 to 0.2 M in deionized water. Indium tin oxide (ITO) coated glass substrates (thickness: 145 ± 10 nm, resistivity: 20 ± 2 Ω/sq) were purchased from Thin Film Devices Inc. All ITO substrates were cleaned through ultrasonic treatment in isopropanol and acetone for 10 minutes and then by $O_2$ ICP/RIE etch for 300 s to remove organic contaminants.

*Preparation of crosslinked PHEMA hydrogel films on ITO substrates.*

Poly(2-hydroxymethyl methacrylate) (PHEMA) hydrogel precursor was synthesized by reacting of 19.7 mM hydroxyethyl methacrylate, 0.3 mM trimethoxysilyl propyl methacrylate (providing crosslinkable functional group) and 0.15 mM azobisisobutyronitrile in 10.5 g dimethylformamide at 55 °C for 3 hours. The PHEMA films with different thickness were prepared on ITO substrates by spin-coating 20 wt% dimethylformamide solution of PHEMA at speeds between 1000 and 5000 rpm for 60 s. The thickness of PHEMA films can be easily controlled by adjusting the concentration of PHEMA solution and spin speed. The PHEMA films were cured on at 100 °C on a hot plate for 6 hours to obtain crosslinked hydrogel films.

*Electrodepositon of ZnO nanostructures*

ZnO thin films with various nanostructures were fabricated by cathodic eletrodeposition from zinc nitrate aqueous solutions. The eletrodeposition and electrochemical measurements were performed with a CHI 600D electrochemical workstation. Scheme 1 shows the schematic diagram for the fabrication of ZnO nanostructures on crosslinked PHEMA coated ITO by electrochemical deposition. A standard three-electrode configuration was used. The crosslinked PHEMA coated ITO was used as the working electrode (cathode), a Pt wire as counter electrode, and an Ag/AgCl (KCl saturated) as the reference electrode. The electrochemical cell was placed in the water bath and the deposition temperature fixed at 70 °C. All experiments were carried out potentiostatically in the range of −1000 to −1200 mV. Immediately after electrodeposition, the ZnO films were removed from the cell and rinsed with deionized water and dried with $N_2$ flow.

*Characterization*

Film thickness measurements of PHEMA and ZnO films were performed with a Veeco Dektak 150 profilometer. A Trion Technology Phantom III inductively coupled plasma (ICP) reactive ion etcher (RIE) was used to clean the ITO substrates. X-ray diffraction (XRD) patterns of ZnO thin films were collected using Cu Kα radiation (λ = 1.5406 Å) on a PANalytical X'Pert PRO diffractometer operating at 45 kV and 40 mA. The surface morphology and structures of as-prepared ZnO films were observed by a JEOL JSM 6320F scanning electron microscopy (SEM) operated at 5 kV. XPS measurements were performed on a Physical Electronics Quantum 2000 Microprobe XPS instrument using a 50 W monochromatic Al X-ray (1486.7 eV) source at a takeoff angle of 45° with a 200 μm spot area. The depth profiles of samples were acquired with 2 keV Ar+ ion sputtering. The XPS data were analyzed using the Multipak software.

Hybrid UV photodetector devices based on Au/ZnO-hydrogel/ITO structures were fabricated. Top Au electrodes were thermally evaporated through shadow masks onto the electrodeposited ZnO layers under vacuum. A 365 nm UV light was used as the light source for the photoconductivity experiments. The I–V characteristics of the devices were measured with a Keithley 4200 semiconductor characterization system at room temperature in ambient air.

## Acknowledgements

The authors acknowledge financial support from the Center for Hierarchical Manufacturing (CHM) at the University of Massachusetts Amherst (CMMI-1025020), the MRSEC on Polymers at the University of Massachusetts Amherst (DMR-0820506), kind support from the Panasonic Boston Research Laboratory. This work was also supported by the Program for Changjiang Scholars and Innovative Research Team in University (No. PCSIRT1126), National Natural Science Foundation of China (no. 20903034 and 11274093) and SRF for ROCS, SEM.

**Supporting Information**

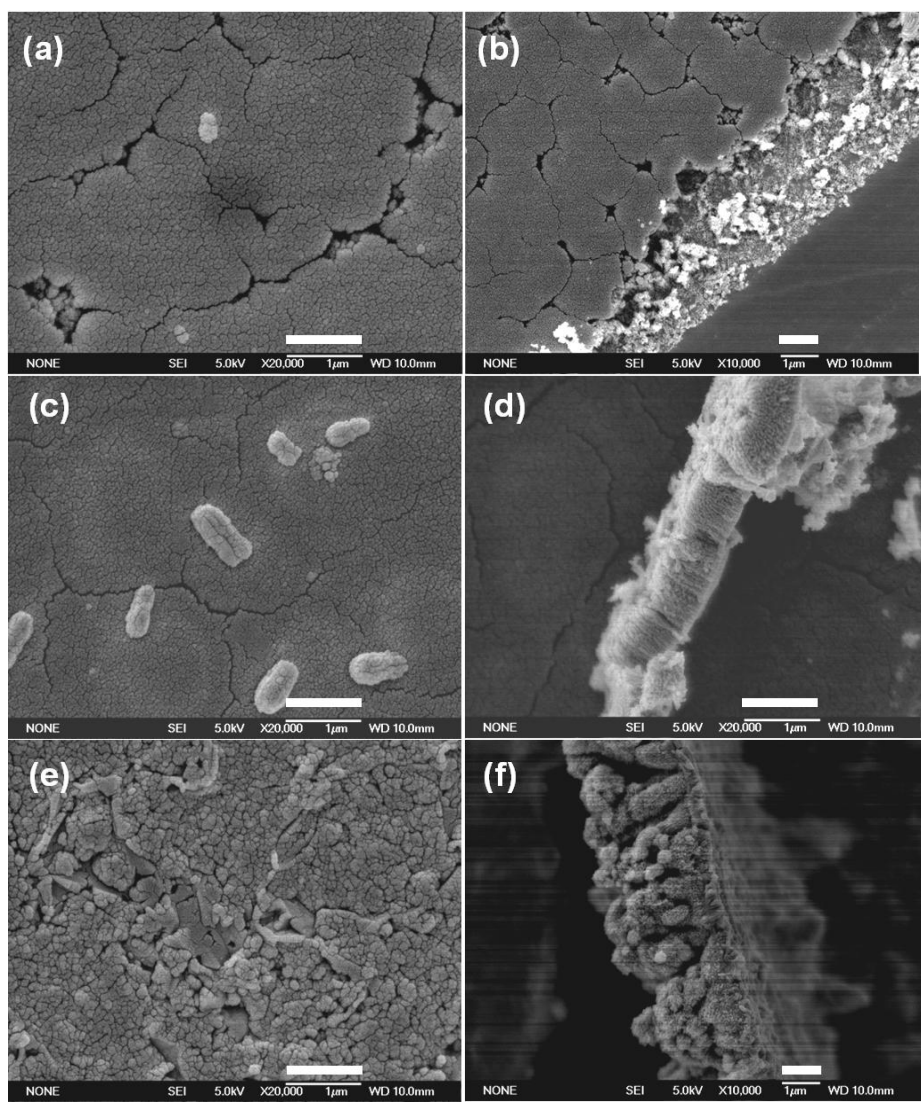

**Figure S1.**

Top-view and cross-section SEM images of ZnO/PHEMA hybrid films calcinated at 500 $^{o}$C for 2h. The ZnO films were obtained by electrodeposition at 70 $^{o}$C under -1.1 V on 500-nm-thick PHEMA hydrogel coated ITO substrates in 0.01 M $Zn(NO_3)_2$ solutions for (a, b) 400 s, (c, d) 600 s and (e, f) 1000 s.

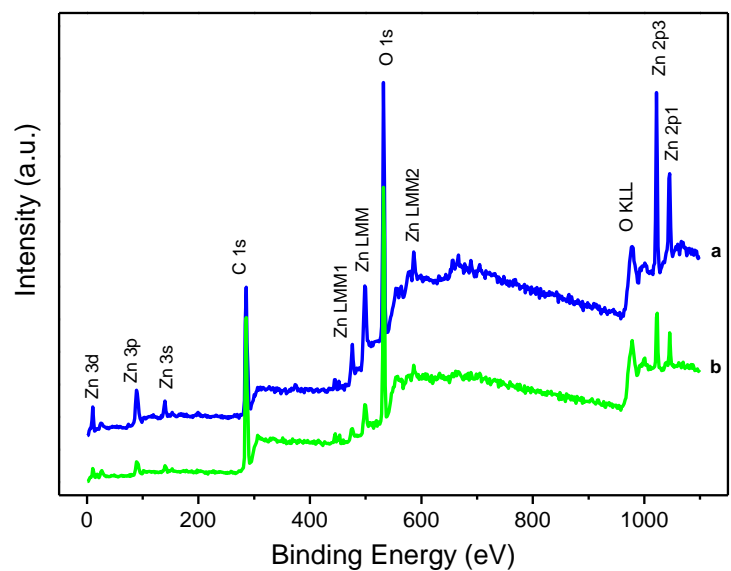

**Figure S2.** XPS survey spectra for the ZnO/PHEMA composite films electrodeposited at 70 °C under -1.1 V on PHEMA hydrogel coated ITO substrates in 0.01 M Zn(NO$_3$)$_2$ solution for (a) 400 s and (b) 1000 s.